\documentclass[dvips,twocolumn,aps,showpacs,amssymb,10pt]{revtex4}
\usepackage{graphicx}% Include figure files
\usepackage{dcolumn}% Align table columns on decimal point
\usepackage{bm}% bold math
\usepackage{epsfig}
\usepackage{color}
\usepackage{longtable}
\usepackage{soul}
\usepackage{amssymb,amsmath}
\usepackage{verbatim}
\everymath=\expandafter{\the\everymath\displaystyle}
\newcommand{\ds}{\displaystyle}

\newcommand{\bra}[1]{\mathinner{\langle{#1}|}}
\newcommand{\ket}[1]{\mathinner{|{#1}\rangle}}
\newcommand{\braket}[2]{\langle #1|#2\rangle}
\newcommand{\redbra}[1]{\langle{#1}|\!|}
\newcommand{\redket}[1]{|\!|{#1}\rangle}

\newcommand{\dd}{\mathrm{d}}

\newcommand{\uvector}[1]{\hat{\bm{\mathbf{#1}}}}
\renewcommand{\vec}[1]{\mathbf{#1}}

\def\etal{et al.}                                        %

\begin{document}
\preprint{}
\title{Relativistic total cross section and angular distribution for Rayleigh scattering by atomic hydrogen}
%
% ------------------------------- Authors ------------------------------------ %
%

\author{L. Safari$^{1,}$\footnote{laleh.safari@oulu.fi}, P. Amaro$^{2,3}$, S. Fritzsche$^{1,4}$, J. P. Santos$^{3}$ and F. Fratini$^{1}$} 

\affiliation{\it
$^1$ Department of Physics, University of Oulu, Box 3000, FI-90014 Oulu, Finland\\
$^2$ Physikalisches Institut, Universit\"{a}t Heidelberg, D-69120 Heidelberg, Germany\\
$^3$ Centro de F\'{i}sica At\'{o}mica, Departamento de F\'{i}sica, Faculdade de Ci\^{e}ncias e Tecnologia, FCT, Universidade Nova de Lisboa, P-2829-516 Caparica, Portugal\\
$^4$ GSI Helmholtzzentrum f\"{u}r Schwerionenforschung, D-64291 Darmstadt, Germany} 
\date{\today \\[0.3cm]}%
%
%
%
%++++------------------------------- Abstract -----------------------------------+++++++++++++++++ %
\begin{abstract}

We study the total cross section and angular distribution in Rayleigh scattering by hydrogen atom in the ground state, within the framework of Dirac relativistic equation and second-order perturbation theory.
The relativistic states used for the calculations are obtained by making use of the finite basis set method and expressed in terms of B-splines and B-polynomials. We pay particular attention to the effects that arise from higher (non-dipole) terms in the expansion of the electron-photon interaction. It is shown that the angular distribution of scattered photons, while it is symmetric with respect to the scattering angle $\theta$=90$^\circ$ within the electric dipole approximation, becomes asymmetric when higher multipoles are taken into account. 
The analytical expression of the angular distribution is parametrized in terms of Legendre polynomials. 
Detailed calculations are performed for photons in the energy range 0.5 to 10 keV. When possible, results are compared with previous calculations.

\end{abstract}

%--------------------------------------------------------

%--------------PACS------------------
\maketitle

%------------------------------- Text ----------------------------------- 
%
%
%
%
%
\section{Introduction}\label{sec:intoduc}

Rayleigh scattering (also called {\it coherent scattering}) denotes the elastic scattering of photons by bound electrons and is usually mentioned in order to explain the blue sky and the red sunset \cite{C.F.Bohren:85,M.Sneep:05}.  

Rayleigh scattering, apart from being a subject of theoretical study by itself, has also interesting applications to different fields. Recently, with the availability of X-Ray polarization detectors as well as synchrotron and FEL sources, new experimental information is emerging in the X-ray regime, and, consequently, the demand for accurate theoretical predictions is arising \cite{P. Duvauchelle:00,N. S. Kampel1:12,D. G. Norris:12}.

The total cross section for Rayleigh scattering has been widely investigated during the past decades, both within relativistic and non-relativistic frameworks \cite{M.Gavrila:67, V. Florescu:90, H.W.Lee:04, H.R.Sadeghpour:92, A.I.Miller:69}.
For instance, in 2007 Nganso and Njock have presented analytical results for Raman and Rayleigh scattering in hydrogenlike ions based on fully relativistic wavefunctions and the Sturmian expansions of the Dirac-Coulomb Green function \cite{H. M. T. Nganso 07}. More recently, Costescu \etal{} have analyzed Rayleigh scattering amplitudes in ions and neutral atoms by using the independent particle model \cite{A. Costescu:2011}.

In contrast to the total cross section, less attention has been paid to the angular distribution of the scattered photons (i.e. the angle-differential cross section) as well as their polarization properties. 
To the best of our knowledge, a fully relativistic treatment of the angular distribution in Rayleigh scattering by hydrogen atom based on Dirac equation has not been performed yet \cite{A. Costescu:94, S.C.Roy:99, S.C.Roy:85, S.C.Roy:93, J.Eichler:85}. Hydrogen atom, although is not, perhaps, the best choice for experimental investigation, is the most abundant element in the Universe and therefore is of great interest in astrophysics. In the extended atmosphere around a giant star, for example, near-ultraviolet photons can be significantly scattered by atomic hydrogen \cite{H. Isliker:89}.

In the present work, we study the total cross section and the angular distribution of the scattered photons in Rayleigh scattering by atomic hydrogen in the ground state. 
The calculations are carried out within a relativistic framework through the use of finite basis sets for the Dirac equation constructed from B-splines and B-polynomials.
The angular distribution is furthermore parametrized in terms of Legendre polynomials and the resulting coefficients are plotted against the photon energy.
The photon energy range we investigate is 0.5 to 10 keV. By comparing with previous works,
good agreement is found in analyzing the total cross section in the whole energy range. 
We show that higher multipoles in the expansion of the electron-photon interaction operator play an important role both in the total cross section and, especially, in the angular distribution of the scattered photons. While, for the total cross section, the non-relativistic dipole approximation is adequate for energies  below $\sim$ 3 keV, in the angular distribution non-dipole effects become important already for photon energies $\gtrsim$ 500 eV.

This article is structured as follows. In section \ref{sec:geometry}, we present the geometry we consider for the scattering process and the notation used. In section \ref{sec:Theory}, the essential theoretical background needed for the calculations is presented together with a detailed explanation of the theoretical quantities that we intend to investigate, namely the total cross section and the angular distribution. In section \ref{sec:Comp}, we explain the numerical method we use in order to obtain the total cross section and the angular distribution. In section \ref{sec:ResultsDiscussion}, we present our results, by showing the impact that high multipoles in the expansion of the electron-photon interaction have.
Finally, a summary is given in section \ref{sec:SumConcl}, together with a few perspectives for further theoretical studies.

\section{Atomic system and geometry}
\label{sec:geometry}

To explore the cross section and the angular distribution in Rayleigh scattering, let us first introduce the atomic system and the geometry under which the distribution of scattered photons is considered. \par
\begin{figure}[h]
\includegraphics[scale=0.8]{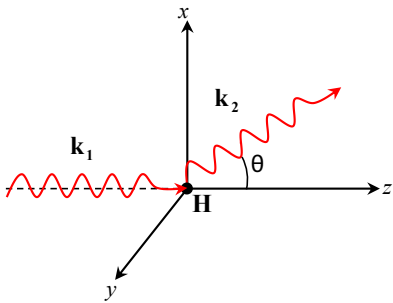}
\caption{(Color online) The adopted geometry for the scattering process is displayed. The scattering angle $\theta$ uniquely defines the direction of the outgoing photon in the scattering ($xz$) plane. The hydrogen atom is placed at the origin of the coordinate axes.}
\label{Rayleigh geometry}
\end{figure}
Our system consists of a hydrogen atom in the ground state which is irradiated by light, as shown in Fig.~\ref{Rayleigh geometry}.
Incident (scattered) light has energy $E_{\gamma_{1 (2)}}= \hbar \omega_{1 (2)}$, propagation vector $\vec{k_{1 (2)}}$ and polarization vector $\uvector{\bm{\epsilon}}_{1 (2)}$, where $\hbar$ is the Planck constant and $\omega$ is the angular frequency of light. Here and in the following, $\uvector{A}$ denotes the unit vector $\vec{A}/|\vec{A}|$, for any vector $\vec{A}$.\\
We adopt the quantization ($z$) axis along the direction  of the incident light ($\vec{k_1}$).
As we will see in Sec. \ref{Multipole decompositions}, such a choice of quantization axis simplifies the multipole expansion of the electron-photon interaction operator.
The scattered light propagates along the direction $\vec{k_2}$ at angle $\theta$ with respect to the $z$ axis. The scattering plane ($xz$) is defined by the incoming and outgoing photon directions.

Now that we have explained the adopted geometry and notation, we are ready to present the theory of Rayleigh scattering.

\section{Theory}\label{sec:Theory}

\subsection{Evaluation of the transition amplitude}
\label{Rayleigh}

There has been a lot of theoretical studies on the calculation of the transition amplitude in Rayleigh scattering both in non-relativistic and relativistic regimes \cite{M.Gavrila:67, V. Florescu:90, H.W.Lee:04, H.R.Sadeghpour:92, A.I.Miller:69, H. M. T. Nganso 07}. In those studies, the analysis of the scattering cross section is usually traced back to the second-order bound-bound transition amplitude.
The procedures to derive such an amplitude are explained in standard textbooks of Quantum Mechanics \cite{Filippo:11, Akhiezer:65}, and are based on second-order time-dependent perturbation theory.
With some straightforward algebraic manipulation, the transition amplitude for Rayleigh (and Raman) scattering can be cast in the form
\begin{eqnarray}
\label{Mamplitude}
           & &  \mathcal{M}^{-\gamma\gamma}(i\to f) = \nonumber\\
           & &  \sum_{\nu}\frac{\bra{f}\bm\alpha\cdot\uvector{\bm{\epsilon}}^*_2\, e^{-i\vec k_2 \vec r}
                \ket{\nu}\bra{\nu}\bm\alpha\cdot\uvector{\bm{\epsilon}}_1\,
                 e^{i\vec k_1 \vec r}\ket{i}}{\omega_{\nu i}-\omega_1} + \nonumber\\
           & &  \sum_{\nu}\frac{\bra{f}\bm\alpha\cdot\uvector{\bm{\epsilon}}_1\, e^{i\vec k_1 \vec r}
                \ket{\nu}\bra{\nu}\bm\alpha\cdot\uvector{\bm{\epsilon}}^*_2\,
                 e^{-i\vec k_2 \vec r}\ket{i}}{\omega_{\nu i}+\omega_2},
\end{eqnarray}
where $\omega_{\nu i} = (E_{\nu}-E_i)/\hbar$ is the transition frequency between states $\ket{\nu}$ and $\ket i$, $\bm\alpha$ is the vector of Dirac matrices and $E_{\nu (i)}$ is the energy of the intermediate (initial) atomic bound state.
Here, the transition operator $\uvector{\bm{\epsilon}}_{1,2} e^{i\vec k_{1,2} \vec r}$ describes the relativistic electron-photon interaction. 
As indicated in Eq.~(\ref{Mamplitude}), the summation over the intermediate states runs over the complete one-particle spectrum $\ket{\nu}$, including a summation over the discrete part of the spectrum as well as the integration over the positive and negative energy continua. 
The initial state $\ket i$ and final state $\ket f$ of atomic hydrogen have well-defined angular momentum $j$, angular momentum projection $m_j$ and parity $(-1)^l$, where $l$ is the orbital angular momentum of the larger component of the Dirac spinor.
In the following, we will rewrite them respectively as $\ket{\beta_i,j_i,m_{j_i}}$ and 
$\ket{\beta_f,j_f,m_{j_f}}$, where $\beta$ is a collective label used to denote all the additional quantum numbers needed to specify the atomic states but for $j$ and $m_j$. In particular, $\beta$ refers to $n$ (principal quantum number) and $l$.

\begin{figure}[h]
\includegraphics[scale=0.5]{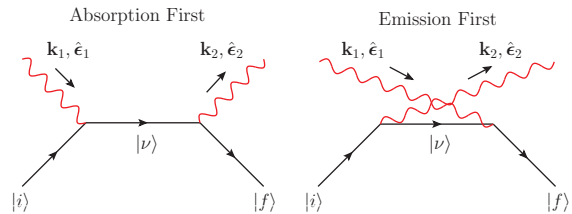}
\caption{(Color online) Feynman diagrams which correspond to the first and second term of the transition amplitude \eqref{Mamplitude}, respectively.}
\label{Feynman diagrams}
\end{figure}
The transition amplitude in Eq.~(\ref{Mamplitude}) can be interpreted in the language of Feynman diagrams from Quantum Electrodynamics theory. The whole scattering process can be described by the two Feynman diagrams shown in Fig.~\ref{Feynman diagrams}, which will be hereinafter called ``absorption first'' and ``emission first'', respectively \cite{P. M. Bergstrom:93}. The middle lines in the diagrams refer to a set of virtual intermediate states that represent both electron and positron states. 
Absorption first diagram corresponds to the first term of the transition amplitude (\ref{Mamplitude}) while emission first diagram corresponds to the second term.

Due to the conservation of energy, the quantities $ E_{\gamma_{1,2}}$ and $ E_{f,i}$ are simply related by the equation
\begin{equation}
\label{conservation of energy}
   E_f-E_i = E_{\gamma_1}-E_{\gamma_2} ~.
\end{equation}
Since Rayleigh scattering is an elastic process, the initial and final states are the same, $\ket{i}\,=\,\ket{f}$, and, thus, Eq.~(\ref{conservation of energy}) simplifies to $ E_{\gamma_1}=E_{\gamma_2}\equiv E_{\gamma}$.

For gaining deeper insights in the transition amplitude (\ref{Mamplitude}), in the next subsection we shall decompose the electron-photon interaction operator in terms of its spherical tensor components.

\subsection{Multipole decompositions of the photon fields}
\label{Multipole decompositions}

Since we want to study the angular properties of the scattered photons, we need to rewrite the vector plane wave $\uvector{\bm{\epsilon}} e^{i\vec k \vec r}$ in terms of elements with well defined angular momentum properties.
This can be done by using the multipole decomposition of the vector plane wave in terms of spherical tensors \cite{M.E. Rose:57}.
Such a decomposition reads
\begin{eqnarray}
\label{multi-pole decomposition final}
               \bm{{\uvector\epsilon}} e^{i\vec k \vec r} &=&
               \sqrt{2\pi}\sum^{+\infty}_{L=1}\sum^{L}_{M=-L}\sum_{p=0,1}  
                  i^L[L]^{1/2} (i\lambda)^p\,\bm{a}^p_{LM}(k,\vec r) \nonumber\\
               & & \times\, D^L_{M\lambda}(\varphi_k,\theta_k,0)    ~,
\end{eqnarray}
where
\begin{eqnarray}
\label{a(p,LM)}
    \bm{a}^p_{LM}(k,\vec r) = \left\{
\begin{array}{ll}      
                      {\vec A}^{(m)}_{LM}(k,\vec r)& \quad \textrm{if } p = 0~,\\[0.4cm]
                      {\vec A}^{(e)}_{LM}(k,\vec r)& \quad \textrm{if } p = 1~,
\end{array} \right.
\end{eqnarray}
with $\lambda$ being the photon helicity. We have defined $[L_1,L_2,...,L_n]=(2L_1+1)(2L_2+1)...(2L_n+1)$ for the sake of simplicity.
Each term $\bm{a}^p_{LM}(k,\vec r)$ has angular momentum $L$, angular momentum projection M and parity $(-1)^{L+1+p}$. 
The standard notation ${\vec A}^{(e,m)}_{LM}$ is used to denote the electric ($e$) and magnetic ($m$) multipole fields. 
Each of these multipoles can be expressed in terms of the spherical Bessel functions $j_L(kr)$ and the vector spherical harmonics ${\vec T}^M_{JL}(\bm{\uvector r})$ of rank $L$ as
\begin{eqnarray}
\label{A(m,e,LM)}
   {\vec A}^{(m)}_{LM}(k,\vec r) &=& j_L(kr) {\vec T}^M_{LL}(\bm{\uvector r})~,  \nonumber\\
   {\vec A}^{(e)}_{LM}(k,\vec r) &=& j_{L-1}(kr)\sqrt{\frac{L+1}{2L+1}}{\vec T}^M_{LL-1}(\bm{\uvector r})
   \nonumber\\                   &-& j_{L+1}(kr)\sqrt{\frac{L}{2L+1}}{\vec T}^M_{LL+1}(\bm{\uvector r}),
\end{eqnarray}
where ${\vec T}^M_{JL}(\bm{\uvector r})$ is defined as
\begin{eqnarray}
\label{T(JLM)}
    & & {\vec T}^M_{JL}(\bm{\uvector r}) = \nonumber\\
    & &  \sum^1_{m=-1}{\braket{L,M-m,1,m}{J,M}\,Y^{M-m}_L}(\varphi_r, \theta_r)\, \uvector\xi_m~. 
\end{eqnarray}
The spin spherical tensor $\bm{\uvector{\xi}}_m$ is defined by
\begin{equation}
\label{spin spherical tensor}
    \bm{\uvector{\xi}}_m = \left\{
\begin{array}{ll}      
                      +\frac{1}{\sqrt{2}}(\uvector{\bm x}-i\uvector{\bm y})& \quad \textrm{if } m = -1 ~,\\
                      \uvector{\bm z}                                     & \quad \textrm{if } m = 0  ~,\\
                     -\frac{1}{\sqrt{2}}(\uvector{\bm x}+i\uvector{\bm y})& \quad \textrm{if } m = +1~.
\end{array} \right.
\end{equation}

As seen from Eq.~(\ref{multi-pole decomposition final}), the angular dependence of the photon emission results from the Wigner (rotation) matrices.
The Wigner matrices transform the multipole fields, with original quantization axis along the photon propagation direction,
into the fields with quantization axis along the 
$\uvector{\bm z} || \vec k_1 $ direction.
Such a choice of quantization axis allows us to describe the second photon direction by means of a single polar angle $\theta$. On account of this, we will consider $D^{L_2}_{M_2\lambda_2}(\bm{\uvector k}_2 \to \bm{\uvector z})= d^{L_2}_{M_2\lambda_2}(\theta)$ and $D^{L_1}_{M_1\lambda_1}(\bm{\uvector k}_1 \to \bm{\uvector z})= \delta_{M_1,\lambda_1}$ in evaluating the transition amplitude (\ref{Mamplitude}).

In the following subsection, by substituting the multipole decomposition (\ref{multi-pole decomposition final}) into the transition amplitude (\ref{Mamplitude}), we shall show how the total cross section and the angular distribution can be obtained in terms of the transition amplitude.

\subsection{Total cross section and angular distribution}

The general form of the relativistic transition amplitude for Rayleigh scattering is described by Eq.~(\ref{Mamplitude}). Such an amplitude contains the whole information on the properties of the scattered radiation.
For instance, the angular distribution function can be
written in terms of the squared transition amplitude as (SI units)
\begin{widetext}
\begin{equation}
\label{dsig/domega}
\begin{split}
 \frac{\dd \sigma}{\dd \Omega}\equiv&\,\frac{1}{2\pi}\frac{\dd\sigma}{\dd\cos\theta}
   =\,\frac{\alpha^2c^2}{2(2j_i+1)}\sum_{\substack{m_{j_i},m_{j_f}\\ \lambda_1,\lambda_2}}\Big|{\mathcal{M}}^{-\gamma\gamma}(i\to f)\Big|^2 ~,
\end{split}
\end{equation}
where $\alpha$ ($\alpha=\frac{e^2}{4\pi\epsilon_0\hbar c}\approx\frac{1}{137}$)
is the electromagnetic coupling constant and $c$ is the speed of light in vacuum. Since we do not investigate polarizations in the present article, in Eq.~\eqref{dsig/domega} we have summed over the final and averaged over the initial atom and photon polarizations.
By combining Eqs.~\eqref{Mamplitude}, \eqref{multi-pole decomposition final} and \eqref{dsig/domega} and
by employing the Wigner-Eckart theorem \cite{Sakurai:94}, the transition amplitude can be written as
\begin{equation}
\label{eq:M}
\begin{array}{lcl}
       \ds\mathcal{M}^{-\gamma\gamma}(i\to f)&=&
        2\pi\sum_{\substack{L_1, L_2\\M_2}}\sum_{p_1p_2}(+i)^{L_1-L_2+p_1+p_2}
        [L_1,L_2]^{1/2}(-1)^{\lambda_2}(\lambda_1)^{p_1}(\lambda_2)^{p_2}
        d^{L_2}_{M_2-\lambda_2}(\theta)
\\[0.8cm]
   &&   \times\;\sum_{j_\nu}
        (-1)^{-j_{\nu}}\frac{1}{(2j_{\nu}+1)^{1/2}}
        \Big(\Theta^{j_{\nu}}(1,2)S^{j_{\nu}}(1,2)
        +\Theta^{j_{\nu}}(2,1)S^{j_{\nu}}(2,1)
        \Big)~.
\end{array}
\end{equation}
The reduced (second-order) matrix element is given by
\begin{equation}
\label{eq:S_J}
S^{j_\nu}(1,2)=
\sum_{\beta_{\nu}}
\frac{\redbra{\beta_i,j_i}\bm\alpha\cdot\bm a_{L_1}^{p_1}(k_1,\vec r) \redket{\beta_{\nu},j_{\nu}}
\redbra{\beta_{\nu},j_{\nu}}\bm\alpha\cdot\bm a_{L_2}^{p_2}(k_2,\vec r)\redket{\beta_{i},j_{i}}}
{\omega_{\nu i}+\omega_2},
\end{equation}
and $S^{j_\nu}(2,1)$ is obtained from \eqref{eq:S_J} by i) interchanging the label 1 with 2 and ii) replacing the positive sign in the denominator with a negative sign. This latter replacement is given by the fact that, while the second photon is emitted, the first photon is absorbed by the atom.\\
Following the notation used in \cite{S.P. Goldman:81}, we have furthermore defined  
\begin{equation}
\label{eq:ThetaS}
\begin{array}{l}
\Theta^{j_{\nu}}(1,2)=\sum_{m_{j_\nu}}(-1)^{m_{j_f}+m_{j_{\nu}}}(2j_{\nu}+1)^{1/2}\left(
\begin{array}{ccc}
j_f&L_1&j_{\nu}\\
-m_{j_f}&\lambda_1&m_{j_{\nu}}
\end{array}
\right)\left(
\begin{array}{ccc}
j_{\nu}&L_2&j_{i}\\
-m_{j_{\nu}}&M_2&m_{j_{i}}
\end{array}
\right)~,
\end{array}
\end{equation}
where $\Theta^{j_{\nu}}(2,1)$ is obtained from Eq.~\eqref{eq:ThetaS} by replacing $L_1\leftrightarrow L_2$ and $\lambda_1\leftrightarrow M_2$.

The transition amplitude \eqref{Mamplitude} formally includes the infinite summation over all the multipoles combinations $p_1L_1\,p_2L_2$, where $p = 1$ (0) refers to $E$ ($M$). Using standard notation, we may write
\begin{equation}
\mathcal{M}^{-\gamma\gamma}(i\to f)\simeq E1E1 + E1M1 + M1M1 + M1E1 + E2E1 + ... ~.
\label{eq:Mexpansion}
\end{equation}
However, according to parity and angular momentum selection rules, the number of allowed (i.e. non-zero) terms in the summation is restricted to certain combinations of the indices $L_1$, $L_2$, $p_1$ and $p_2$. 
Since the atomic (initial and final) state that we consider for the calculations is $1s_{1/2}$ (the ground state), the allowed multipole terms are the ones with equal parity and $|L_1-L_2|\leq 1$. More explicitly, the multipoles that must be considered are $E1E1$, $M1M1$, $E1M2$, $M2E1$, ... .\\
Finally, integrating \eqref{dsig/domega} over the second photon directions, we get the total cross section as (SI units)
\begin{equation}
\label{sigma}
\sigma = 
       \frac{\alpha^2c^2}{2(2j_i+1)}\sum_{\substack{m_{j_i},m_{j_f}\\ \lambda_1,\lambda_2}}
       \sum_{L_2M_2}\Big|\tilde{\mathcal{M}}^{-\gamma\gamma}(i\to f)\Big|^2~,
\end{equation}
where
\begin{equation}
\label{Mtild}
\begin{array}{lcl}
\tilde{\mathcal{M}}^{-\gamma\gamma}(i\to f)
&=& 4\pi\sqrt{\pi}\sum_{L_1}\sum_{p_1p_2}(+i)^{L_1+p_1+p_2}[L_1]^{1/2}
    (\lambda_1)^{p_1}(\lambda_2)^{p_2}
\\[0.6cm]
&&   \times\;\sum_{j_\nu}
    (-1)^{-j_{\nu}}\frac{1}{(2j_{\nu}+1)^{1/2}}
    \Big(\Theta^{j_{\nu}}(1,2)S^{j_{\nu}}(1,2)
    +\Theta^{j_{\nu}}(2,1)S^{j_{\nu}}(2,1)\Big)~.
\end{array}
\end{equation}
\end{widetext}
Comparing Eqs. \eqref{sigma} and \eqref{dsig/domega}, we note that the summation over the multipoles of the outgoing photon ($L_2$, $M_2$), while is inside the modulus squared in the angular distribution, comes outside the modulus squared in the total cross section. By using the notation introduced in Eq.~\eqref{eq:Mexpansion}, we may better underline the consequences of this fact: Interference terms of the type ($E1E1$)($M1M1$)$^*$ or ($M1M1$)($E1E1$)$^*$ are forbidden in the calculation of the total cross section, while they are allowed in the calculation of the angular distribution. Thus, higher (non-dipole) multipoles will start giving non-zero contributions in the angular distribution at {\it lower} photon energies than in the total cross section. This effect will be analyzed in detail in Sec. \ref{sec:ResultsDiscussion}, where 
expressions \eqref{dsig/domega} and \eqref{Mtild} will be used to explore the total cross section as well as the angular distribution in the Rayleigh scattering by hydrogen atom. However, prior to doing that, we shall explain how the reduced matrix elements (\ref{eq:S_J}) are calculated in the present work. 

%++++++++++++++++++++++++++++++++++++++++++++++++++++++++++++++++++++++++++++++++++++++++++++++++++++
%
%
%
%
%
%
\section{Computation}\label{sec:Comp}

During the last decade, various methods have been investigated for calculating the reduced second order amplitude (\ref{eq:S_J}) as well as the transition amplitude (\ref{Mamplitude}) \cite{H. M. T. Nganso 07, A. Costescu:2011, A. Costescu:2007, A. Costescu:94, F. Fratini:11, A. Surzhykov:10} . 
In practice, of course, the summation over the complete spectrum contained in \eqref{eq:S_J} is difficult to be performed explicitly. Several approaches and approximation techniques have been proposed to perform such a summation. Among them, the Coulomb-Green function approach has been widely used for investigating both decay of and scattering by atoms and ions \cite{skf2005, M. Alfred:98}.

An alternative approach is the \textit{finite basis set} method \cite{saj1996, jbs1998,D. D. Bhatta:06, ZatBar2008, FroZat2009, A. Surzhykov:11}.
The finite basis set method is based on the supposition that the ion (or atom) is enclosed in a finite cavity with a 
radius $R$ large enough to get a good approximation for the wave functions. Such a restriction leads to a  ``discretized'' continuum part of the atomic or ionic spectrum, and hence to a representation of the Dirac wavefunctions in terms of pseudo basis set functions. This basis set forms a complete set of orthonormal functions \cite{jbs1998}.

In the present work, we calculate the transition amplitude (\ref{eq:S_J}) by using B-splines and B-polynomials as finite basis sets. The B-splines are one of the most commonly used family of piecewise polynomials, since they are well adapted to numerical tasks \cite{jbs1998}. The B-polynomials, or the Bernestein polynomials \cite{D. D. Bhatta:06}, are a good alternative to the B-splines since they allow for analytical finite basis-set calculations. These are polynomial functions of $nth$ degree that are used to obtain the solution of some linear and nonlinear differential equations \cite{D. D. Bhatta:06}. The details of these basis sets, as well as a comparison between them, can be found in Ref. \cite{asfis2011}. Thus, we restrict ourselves to describe the characteristic parameters used in this work. \\
The parameters of the B-splines basis set are the radius of the cavity ($R_{\rm{bs}}$), the number of B-splines ($n_{\rm{bs}}$) and their degree ($k$). As for the B-polynomials, the parameters are the radius of the cavity ($R_{\rm{bp}}$) and the number of B-polynomials ($n_{\rm{bp}}$) (the degree of the B-polynomials is $n_{\rm{bp}}-1$).

The parameters used in both basis sets were optimized in order to obtain stability and agreement of six digits between the results of both basis sets. The optimal parameters are:  $R_{\rm{bs}}=60$~a.u., $n_{\rm{bs}}=60$, $k=9$, $R_{\rm{bp}}=50$~a.u. and $n_{\rm{bp}}=40$. Such set of parameters was already obtained for the case of two photon emission in Refs.~\cite{aspsi2009,asfis2011}.

The finite basis set method has been widely used during the past years to explore the two-photon decay of the metastable $2s_{1/2}$ state in heavy hydrogenlike ions \cite{S.P. Goldman:81,sfi1998}. In contrast to the two photon decay process, to our knowledge no one has ever applied this method to calculate the transition rate for Rayleigh scattering.
\vspace{1.5 cm}

\section{Results and discussion}
\label{sec:ResultsDiscussion}

\begin{figure}[b!]
\includegraphics[scale=0.7]{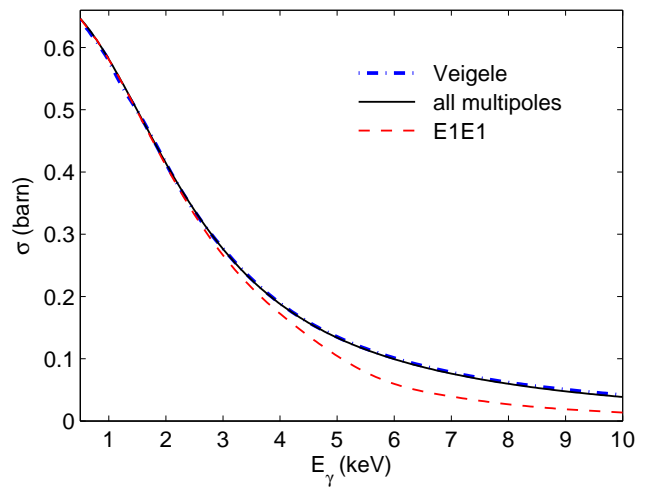}
\caption{(Color online) Total cross section in Rayleigh scattering by atomic hydrogen, as a function of the photon energy. Calculations obtained within the electric dipole approximation (red-dashed line) are compared with those including all of the allowed multipoles (black-solid line) and with those from Veigele \cite{Vei1973} (blue-dot-dashed line).}
\label{fig:Fig3}
\end{figure}
\begin{figure*}[t]
\includegraphics[scale=0.95]{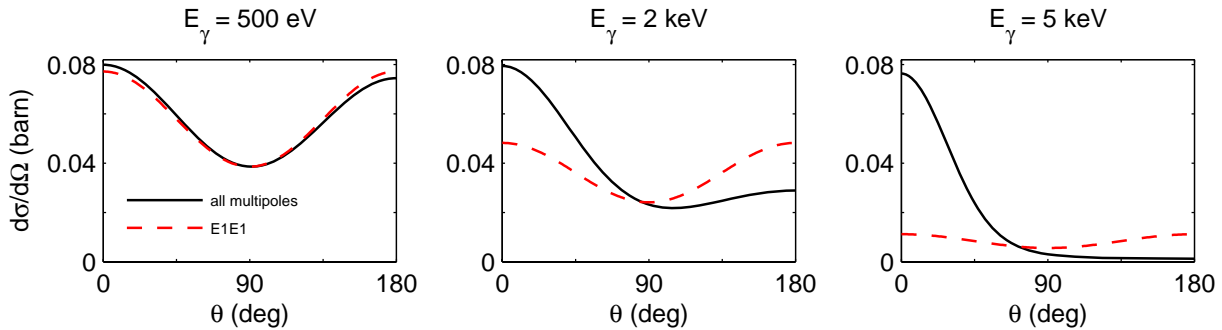}
\caption{(Color online) Angular distribution in Rayleigh scattering by atomic hydrogen in the ground state as a function of the scattering angle $\theta$. Results are calculated within the exact relativistic theory (solid-black line) and within the electric dipole approximation (red-dashed line), for three selected photon energies.}
\label{fig:Fig4}
\end{figure*}
With the formalism developed above, we can now adequately
analyze the total cross section and the angular distribution in Rayleigh scattering by hydrogen atom. The initial and final state considered for the calculations is $1s_{1/2}$.

In Fig.~\ref{fig:Fig3}, we plot the total cross section in the photon energy range 0.5 to 10 keV as given by Eq.~(\ref{sigma}). Calculations obtained within the electric dipole approximation ($E1E1$) and by including all multipoles are separately displayed. Furthermore, results from Veigele's work \cite{Vei1973} have been interpolated and are also shown. Our cross section practically coincides with Veigele's, indicating that there is good agreement between the two calculations. Moreover, it can be seen that for high photon energies the electric dipole approximation deviates from the results of Veigele and of us obtained with the account of all multipoles. For photon energy $\approx$ 3 keV, the electric dipole approximation underestimates the cross section by about 3\%, while, for photon energy $\gtrsim$ 6 keV, it is lower by a factor of two or more. Indeed, this result is not unexpected: in low photon energy range, the first (dipole) term $E1E1$ dominates the transition amplitude, while, for higher photon energy (such that $kr\gtrsim 1$, where $r$ is the atomic radius), higher multipoles play an important role, as it is evident from Eqs. \eqref{multi-pole decomposition final} and \eqref{A(m,e,LM)}.
\begin{table}[b!]
\begin{center}
\caption{Different calculations for the total cross section for Rayleigh scattering by hydrogen in the ground state, for selected values of the photon energy. The results are compared with Veigele \cite{Vei1973} and NIST \cite{Nist}.} \vspace{0.6 cm}
\label{Tab:SelectedRes}
\begin{tabular}{lccc}
\hline\hline \\[-0.1cm] 
$E_{\gamma}$ (eV) $\,\,\,$ & $\,\,$Veigele$\,\,$ & $\,\,$NIST$\,\,$ & $\,\,$This calculation$\,\,$ \\ \hline
500    & 0.642 & n.a.    & 0.6465  \\
1000   & 0.579 & 0.5805  & 0.5812  \\
2000   & 0.414 & 0.4141  & 0.4142  \\
3000   & 0.277 & 0.2760  & 0.2764  \\
5000   & 0.135 & 0.1341  & 0.1341  \\
8000   &0.0618 & 0.0612  & 0.0609  \\
10000  &0.0416 & 0.04121 & 0.04030 \\ \hline
\hline
\end{tabular}\end{center}
\end{table}%
For a detailed comparison, our results for the total cross section are presented in Table \ref{Tab:SelectedRes} together with Veigele's \cite{Vei1973} and NIST's (National Institute of Standards and Technology) \cite{Nist}, for selected photon energy values.

In addition to the total cross section, we analyze the angular distribution of the scattered photons. 
In Fig.~\ref{fig:Fig4}, we display the angular distribution as obtained from Eq.~(\ref{dsig/domega}), for the three photon energy values $E_{\gamma}$= 500 eV, 2 keV  and 5 keV. Calculations have been performed within the electric dipole approximation and by taking into account all multipoles that have a contribution greater than 1 percent. As seen from the figure, the angular distribution for low photon energy values is well described by the electric-dipole approximation and by the expression 
$\sim 1+\cos^2\theta$ \cite{M.Gavrila:67}, as expected from the non-relativistic theory. For photon energy values $\approx$ 2 keV, forward scattering becomes dominant, compared to backward scattering, and leads to an asymmetric angular distribution. For even higher photon energy values ($\gtrsim$ 5 keV), the backward scattering is strongly suppressed and $\approx$ 90\% of scattered photons are found within the interval $0\leq\theta\lesssim60^\circ$. This effect is known from previous calculations and experiments on other ions and atoms. See for instance Ref.~\cite{V. Florescu:90}.

By comparing Figs. \ref{fig:Fig3} and \ref{fig:Fig4}, we notice that the angular distribution is much more sensitive to non-dipole effects than the total cross section: While, for the total cross section, non-dipole effects are suppressed below $\sim$ 3 keV, in the angular distribution they become already important for photon energies $\gtrsim$ 500 eV. As discussed previously (see Eq.~\eqref{Mtild} and the paragraph afterwards), this is a direct consequence of the different structure of the amplitudes $\mathcal{M}$ and $\tilde{\mathcal{M}}$ displayed in Eqs.~\eqref{eq:M} and \eqref{Mtild}, respectively. In particular, following the notation used in Eq.~\eqref{eq:Mexpansion}, the main contribution for the asymmetric shape of the angular distribution comes from the ($E1E1$)($M1M1$)$^*$ and ($M1M1$)($E1E1$)$^*$ terms, which are forbidden in the calculation of the total cross section.

Figure \ref{fig:Fig4} displays the angular distribution for just three specific energy values. To present the angular distribution in the whole energy range, we make an expansion of $\dd \sigma/\dd \Omega$ in terms of Legendre polynomials and we normalize it to the total cross section $\sigma$:
\begin{equation}
\label{eq:parametrization}
\frac{d\sigma}{d\Omega} = \sigma \Big(\beta_0 P_0(\cos \theta)  + \beta_1 P_1(\cos\theta)  + ...\Big) ~,
\end{equation}
where $\beta_i$ are real numbers called \textit{anisotropy  coefficients} \cite{V. Balashov:00}.

\begin{figure}[t!]
\includegraphics[scale=0.7]{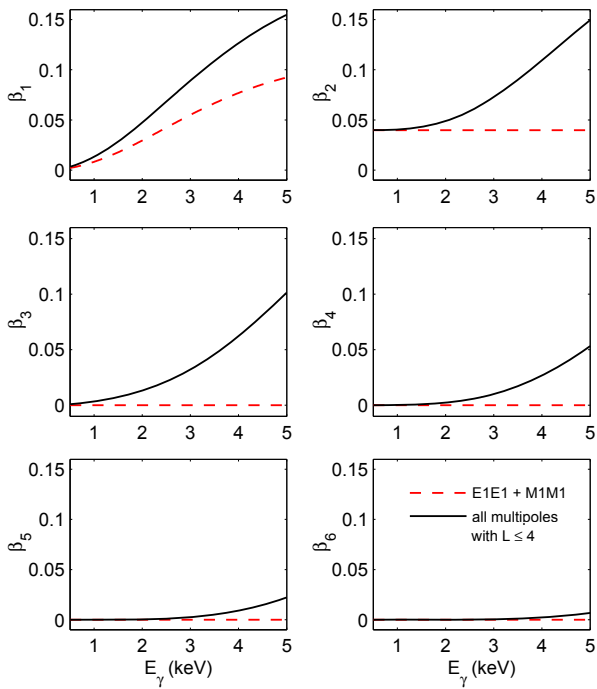}
\caption{(Color online) The anisotropy coefficients $\beta_i$ of Eq. \eqref{eq:parametrization} are plotted against the photon energy. Results obtained within the electric and magnetic dipole approximation (red-dashed line) are compared with results obtained by taking into account all photon multipoles with $L_1\leq4$ and $L_2\leq4$ (black-solid line), which are dipole, quadrupole, octopole and hexadecapole moments of each photon field, both of magnetic and electric type.}
\label{fig:Fig5}
\end{figure}

The results for the $\beta$ coefficients are displayed in Fig.~\ref{fig:Fig5}, within the considered energy range. The first coefficient $\beta_0$ is constant for any photon energy and therefore is not displayed; $\beta_0\approx 0.0796$. The red-dashed curves are obtained by including only the electric and magnetic dipole contributions ($E1E1$+$M1M1$) while the black-solid curves are obtained by taking into account the multipoles with $L_1\leq4$ and $L_2\leq4$. Those multipoles are nothing but dipole, quadrupole, octopole and hexadecapole moments of each photon field, both of electric and magnetic type. From the figure, one immediately notices that, as expected, the lower the photon energy is, the fewer coefficients are needed for an adequate description of the angular distribution.

As seen from Fig.~\ref{fig:Fig5}, for low photon energies we have $\beta_0\approx 2 \beta_2$ and $\beta_i\approx0$ for $i\neq0,2$. By explicitly writing the definitions of Legendre polynomials, one can easily check that the well-known shape $\sim 1+ \cos^2\theta$ for the angular distribution is obtained within the (non-relativistic) Thomson limit.
In contrast, for higher photon energies, higher-order Legendre polynomials must be invoked to well describe the shape of the angular distribution, as the number of multipoles that contribute to it increases. 

Although the calculations have been performed for hydrogen atom, the behavior of the angular distribution for hydrogenlike ions can be qualitatively predicted, at least for photon energies far from the ionization threshold. The angular part of the Dirac wave function remains unchanged when the atomic number $Z$ is increased, while the radial part gets \textit{contracted}: The mean radius, $r$, of the electronic ground state gets shrunk and scales like $r\simeq r_0/Z$, where $r_0$ denotes the radius of hydrogen atom \cite{Bransden:83}. The transition operator in equation \eqref{A(m,e,LM)} contains both the angular operator $\bm{\uvector r}$ and the radial operator $kr$. Thus, while the action of the former will be the same for hydrogenlike ions as for hydrogen atom, the action of the latter will scale as $kr\simeq k r_0/Z=(k/Z) r_0$. Eventually therefore, for energies far from the ionization threshold, the whole angular distribution will scale in the energy domain with a factor of $1/Z$. More explicitly, we approximately obtain the angular distribution of photons which scatter by hydrogenlike atoms with atomic number $Z$ by taking the angular distribution for hydrogen atom and replacing $k\to k/Z$.

\section{Summary and outlook}
\label{sec:SumConcl}

In summary, we studied the total cross section and angular distribution in the Rayleigh scattering of photons by hydrogen atom, based on Dirac relativistic equation and second-order perturbation theory.
 We decomposed the transition amplitude in terms of spherical tensors and (radial) reduced amplitudes, where these latter include a summation over the whole atomic spectrum. This summation was performed by means of finite basis set method. The multipole expansion of the photon fields allowed us to investigate non-dipole effects both in the total cross section and in the angular distribution.
As for the total cross section, we compared our results with previous calculations and good agreement has been found. We then studied the angular distribution of photons in the energy range 500 eV to 5 keV.  
We here found that non-dipole effects become important at much lower energies than for the total cross section. As discussed, this has to be attributed to the different structure of the transition amplitude for angular distribution and total cross section.
Finally, in order to give a more practical account of the angular distribution function, we expanded it with Legendre polynomials and we plotted the resulting coefficients against the photon energy, noticing an evident convergence. 

Owing to recent advances in detector technology \cite{S. Tashenove:06}, photon polarization studies in atomic processes have lately become important \cite{N.L.Manakov:00, F.Fratini:11, F.Fratini:11_2}. Further studies are in progress to investigate the polarization properties of the scattered photons. \vspace{4cm}

\section{acknowledgment}
L. S. and F. F. gratefully acknowledge Helena Aksela for her support. 
P. A. acknowledges the support of German Research Foundation (DFG) within the Emmy Noether program under Contract No. TA 740 1-1. S. F. acknowledges support by the FiDiPro program of the Finnish Academy. J. P. S. and P. A. acknowledge the support by FCT -- Funda\c{c}\~ao para a Ci\^encia e a Tecnologia (Portugal), through the Projects No. PEstOE/FIS/UI0303/2011 and PTDC/FIS/117606/2010, financed by the European Community Fund FEDER through the COMPETE -- Competitiveness Factors Operational Programme.

%
%
%
%                                         bibliography 
% 
%
%

\end{document}